\documentclass[aip,
preprint,]{revtex4-1}
\usepackage{accents}
\usepackage{xcolor}
\usepackage{subfigure}
\usepackage{graphicx}
\usepackage{dcolumn}
\usepackage{bm}
\usepackage{appendix}
\usepackage{color}
\usepackage{float}
\usepackage[colorinlistoftodos]{todonotes}
\usepackage{comment}

\begin{document}

\title{On quasineutral plasma flow in the magnetic nozzle}
\author{A. I. Smolyakov}\altaffiliation{Corresponding Author}
\email{andrei.smolyakov@usask.ca}
\author{A. Sabo}
\affiliation{ University of Saskatchewan, Saskatchewan, Saskatoon SK S7N 5E2, Canada
}
\author{P. Yushmanov}
\author{S. Putvinskii}
\affiliation{ TAE Technologies, 19631 Pauling, Foothill Ranch, CA, 92610, United States
}
\begin{abstract}
Exact solutions for quasineutral plasma acceleration of magnetized plasma in
the paraxial magnetic nozzle are obtained. It is shown that the
non-monotonic magnetic field with a local maximum of the magnetic field is
a necessary condition for the formation of the quasineutral accelerating
potential structure. A global nature of the accelerating potential that
occurs as a result of the constraint due to the regularity condition at
the sonic point is emphasized and properties of such solutions are discussed for the case of general polytropic equation of state for
electrons.
\end{abstract}

\maketitle


\affiliation{University of Saskatchewan, Saskatchewan, Saskatoon SK S7N 5E2, Canada
}

\affiliation{ University of Saskatchewan, Saskatchewan, Saskatoon SK S7N 5E2, Canada
}

\affiliation{ TAE Technologies, 19631 Pauling, Foothill Ranch, CA, 92610, United States
}

\affiliation{ TAE Technologies, 19631 Pauling, Foothill Ranch, CA, 92610, United States
}%


\affiliation{University of Saskatchewan, Saskatchewan, Saskatoon SK S7N 5E2, Canada
}

\affiliation{ University of Saskatchewan, Saskatchewan, Saskatoon SK S7N 5E2, Canada
}

\affiliation{ TAE Technologies, 19631 Pauling, Foothill Ranch, CA, 92610, United States
}

\affiliation{ TAE Technologies, 19631 Pauling, Foothill Ranch, CA, 92610, United States
}%


\affiliation{ University of Saskatchewan, Saskatchewan, Saskatoon SK S7N 5E2, Canada
}

\affiliation{University of Saskatchewan, Saskatchewan, Saskatoon SK S7N 5E2, Canada
}%



%
Plasma flow and acceleration in the magnetic mirror configuration (magnetic
nozzle) is important for many applications such as the expanding magnetic
divertors in fusion applications \cite%
{RyutovAIP2016,GhendrihPPCF2011,TogoNME2019} as well as devices for space
propulsion \cite{WilliamsAIP2003,CollardPSST2019,BoswellPoP2004}. Though
plasma acceleration in the magnetic nozzle was experimentally demonstrated
long time ago \cite{AndersenPF1969}, the exact physical mechanism(s) of the
acceleration  are still actively studied   theoretically
and experimentally.  

Conditions for the formation of the accelerating potential structure have been discussed widely \cite{ChenPoP2006,CohenPoP2003,LongmierPSST2011,GhendrihPPCF2011,TogoNF2019,TakahashiPRL2018,ChenPRE2020}. It is generally understood that the diverging magnetic field is required, however
other processes, such as deviations from quasineutrality  \cite{ChenPoP2006,FruchtmanPRL2006,PerkinsPRL1981,AhedoPRL2009}; electron kinetic, trapping and non-stationary phenomena \cite{RamosPoP2018,AhedoPSST2020,ArefievPoP2009,ArefievPoP2008}; additional geometrical  effects \cite{BennetPSST2018} have also been invoked to explain formation of the accelerating potential.  
 In part, the mechanism of the acceleration in the magnetic nozzle is
obscured by the  relation to a more general problem of double layers (DL). Originally, a double layer  was identified as a localized non-quasineutral
step-like potential structure with the width of the order of the Debye length. Typically, this is 
 a kinetic problem and requires presence of several particle species with different energies \cite{PerkinsPRL1981}. 
Such current-free structures are thought to be
important for particle acceleration in space plasmas. Double layer type structures  were observed in expanding plasma \cite{CharlesAPL2003,HairapetianPRL1990} that lead to the explanations \cite{LiebermanPRL2006,ChenPoP2006,AhedoPRL2009} that involve  group of particles with different energies, non-quasineutral effects,  plasma expansion, and  trapped particles.    In this Letter, we discuss only the quasineutral situation when the accelerating potential is formed by the
magnetic mirror, with the length determined by the  width of the barrier, and typically much wider than the Debye length. For such
structures, the Debye length is not a relevant parameter, and to avoid confusion, here we do not call the quasineutral accelerating layer as a double layer, reserving the latter term for strictly non-quasineutral Debye length scale potential structures.

It is well known that in quasineutral approximation accelerating plasma flow
has a singularity at the sonic point, where the plasma velocity $v_{i}$ is
equal to the local sound velocity $c_{s}$.  A regular smooth solution across the whole
acceleration region from $v_{i}<c_{s}$ to the region with $v_{i}>c_{s}$  by imposing  the regularity condition at 
the sonic point \cite{AhedoPoP2001,CohenZurPoP2002}. As
an example, in case of the ion acceleration in the Hall thrusters,  the
regular solution is obtained by imposing an analytical condition on the ion flux,
electron current and ionization source, eventually, defining the operational
diagram for the Hall thruster discharge \cite{SmolyakovAIAA2019,RomadanovPPR2020}. 

The  magnetic nozzle with the converging-diverging  magnetic field   offers a simplest case of quasineutral plasma acceleration from sub-sonic to supersonic velocity \cite{ManheimerIEEE2001}. An exact solution for a special case of the magnetic field and plasma
parameters profile and isothermal electrons was obtained in Ref. \onlinecite{FruchtmanPoP2012}.  The goal of this letter is to
present exact solutions and discuss their properties for the case of arbitrary profile of the magnetic field and for general
polytropic equation of state for electrons.

The exact solutions presented in this letter clearly demonstrate two related properties of the quasineutral plasma acceleration: (1) a magnetic barrier with the maximum of
the magnetic field is required for the formation of the quasineutral accelerating potential structure; (2) the resulting velocity profiles are "stiff". The
latter means that the velocity profile has no free parameters to match with
the plasma source region, raising interesting questions on how such
solutions can be matched to plasma sources where plasma flow is generated. Despite of their simplicity, these results are not widely appreciated. Many theoretical works deal with plasma acceleration in the region with $%
v_{i}\geq c_{s}$, avoiding the singular point $%
v_{i}=c_{s} $, and  without a discussion how the plasma velocity approaches this point. Full numerical solutions have been obtained in high fidelity two-dimensional models of plasma acceleration in the magnetic nozzle, however the constraints  imposed  by the sonic point transition are not not well studied. As we discuss below, the physics of the singular point defines the global smooth solution for plasma velocity in the whole region from sub-sonic to super-sonic velocity.

We consider a standard paraxial model for stationary quasineutral flow
of plasma with cold magnetized ions. Then plasma flow along the magnetic
field is described by the equations
\begin{equation}
    \nabla _{\Vert }(nV_{\parallel }/B)=0,  \label{continuity2}
\end{equation}
 \begin{equation}
m_inV_{\Vert }\nabla _{\Vert }V_{\Vert }=enE_{\Vert },  \label{energy}
\end{equation}
\begin{equation}
0=en\nabla _{\Vert }\phi -\nabla _{\Vert }p_{e},  \label{potential-gradient}
\end{equation}%
where $V_{\Vert }=\mathbf{V\cdot B/}B$ is the ion velocity along the
magnetic field, and $\nabla _{\Vert }=\mathbf{B\cdot \nabla /}B$ is the
gradient operator along the magnetic field, $p_{e}=n T_{e}$ is the
electron pressure In the paraxial approximation, near the axis of the
slender plasma tube confined by the magnetic field, one can take $\nabla
_{\Vert }$=$\partial /\partial z.$ One has to note that in the paraxial approximation 
two-dimensional effects are included to the first order of the parameter $r/a<1$, where $r$ is the radial distance from the axis, 
and $a$ is the characteristic radial length scale.  It   should be noted that for magnetized plasma flow, when the plasma velocity is strictly along the magnetic field line,  the magnetic field $B$ plays the role of the physical nozzle of the variable cross section, $ \pi r^2 \rightarrow B^{-1}$. Therefore,  many conclusions of this paper will apply to the flow constricted by the physical nozzle.

Assuming the isothermal electrons with $%
T_{e}=const$, \ equations (\ref{continuity2}-\ref{potential-gradient}) are
readily reduced  to a single equation for the ion velocity in the form
\begin{equation}
\left( M^{2}-1\right) \frac{\partial M}{\partial z}=-M\frac{\partial \ln B}{%
\partial z},  \label{gradient-of-M}
\end{equation}%
where the velocity $V_{\parallel }$ is normalized to the speed of sound $%
c_{s},$ $M=V_{\parallel }/c_{s},\,c_{s}^{2}=T_{e}/m_{i}$, which is constant
and uniform for isothermal case. Equation (\ref{gradient-of-M}) is equivalent to the equation 
for the velocity in the Laval nozzle, and also appears in the problem of plasma acceleration by the magnetic pressure \cite{FruchtmanPoP2003} such as in the Magneto-Plasma-Dynamics thrusters.   

Equation (\ref{gradient-of-M}) exhibits the ion-sound point singularity at $%
M=1$ which is a well known feature for the plasma flow in the quasineutral
approximation.  One can see that in the subsonic regime $M<1$, before the
ion-sound point, the plasma is accelerated "kinematicaly" $V_{\Vert
}^{^{\prime }}/V_{\Vert }\simeq B^{^{\prime }}/B>0$, so that  the ion 
acceleration is  mostly due to the decrease of the effective cross-section in the
converging magnetic field and the contribution of the ion inertia to the
acceleration can be neglected. It also means that the variations of plasma
density are also small, as it is expected for subsonic regimes with $V_{\Vert }\ll
c_{s}$. In the supersonic regime, $M>1$, ions continue to be accelerated by
the electric field created by the electron pressure of plasma expanding in
the diverging magnetic field: $m_{i}V_{\Vert }V_{\Vert }^{^{\prime }}=-e\phi
^{^{\prime }}=-T_{e}n^{^{\prime }}/n$ and $n^{^{\prime }}/n\simeq
B^{^{\prime }}/B<0$. The whole acceleration process in the
converging-diverging magnetic field is similar to the gas acceleration in
Laval nozzle \cite{AndersenPF1969}.

It follows from (\ref{gradient-of-M}) that the existence of a regular smooth
solution for $M=M\left( z\right) $ in the whole range from the low velocity $%
M<1$ to the region $M>1$ requires the condition $\partial \ln B\left(
z\right) /\partial z=0$ at the point $M=1$. \ This condition fixes the value
of the derivative of the velocity near the sonic point $M=1$. Expanding the
equation (\ref{gradient-of-M}) near $M=1$ one finds the equation%
\begin{equation}
\left( \frac{\partial M}{\partial z}\right) ^{2}=-\frac{1}{2}\frac{\partial
^{2}\ln B}{\partial z^{2}}>0.  \label{2nd}
\end{equation}%
Therefore, the condition $\partial ^{2}\ln B\left( z\right) /\partial
z^{2}<0 $ at $\ M=1$ is required for the existence of the regular solution,
and the magnetic field should have a maximum at the point where $\partial
\ln B\left( z\right) /\partial z=0$ and $M=1$. In other words, the magnetic
barrier is required for the existence of the regular potential structure
that can accelerate plasma to supersonic velocities.

Equation (\ref{gradient-of-M}) can be integrated giving\cite{ManheimerIEEE2001,FruchtmanPoP2012}
\begin{equation}
\frac{M^{2}}{2}-\frac{1}{2}=\ln \left( M\frac{B_{m}}{B\left( z)\right) }%
\right),  \label{meq}
\end{equation}%
The integration constant (corresponding to the condition (\ref{2nd})) was
chosen to remove the ion sound point singularity at $M=1$, the point where the 
magnetic field has a maximum, $B(z)=B_{m}$.  

Ref. \onlinecite{FruchtmanPoP2012} has provided the particular solutions of equation (\ref{meq}) for the specific
magnetic field profile. Here, we present the general solution for an
arbitrary magnetic field profile. Writing equation (\ref{meq}) in the form
\begin{equation}
M^{2}=\ln \left[ \frac{eM^{2}B_{m}^{2}}{B^{2}\left( z\right) }\right] ,
\label{mglobal}
\end{equation}%
plasma velocity for arbitrary magnetic field profile can be presented in the
form of Lambert function
\begin{equation}
M\left( z\right) =\left[ -W(-b^{2}\left( z\right) /e\right] ^{1/2},
\label{m}
\end{equation}%
Here $W\left( y\right) $ is\ Lambert function, which is the solution of the
equation $W\exp \left( W\right) =y$, $b\left( z\right) \equiv B\left(
z\right) /B_{m}<1$, $e$ is the Euler's number.  It is interesting to note that Lambert function \cite{CorlessACM1996} appears in many physics and applied mathematics applications, including the stationary plasma balance models taking into account neutral dynamics \cite{RaimbaultPoP2007}. 

\begin{figure}[ht]
\begin{center}
\includegraphics[width=120mm]{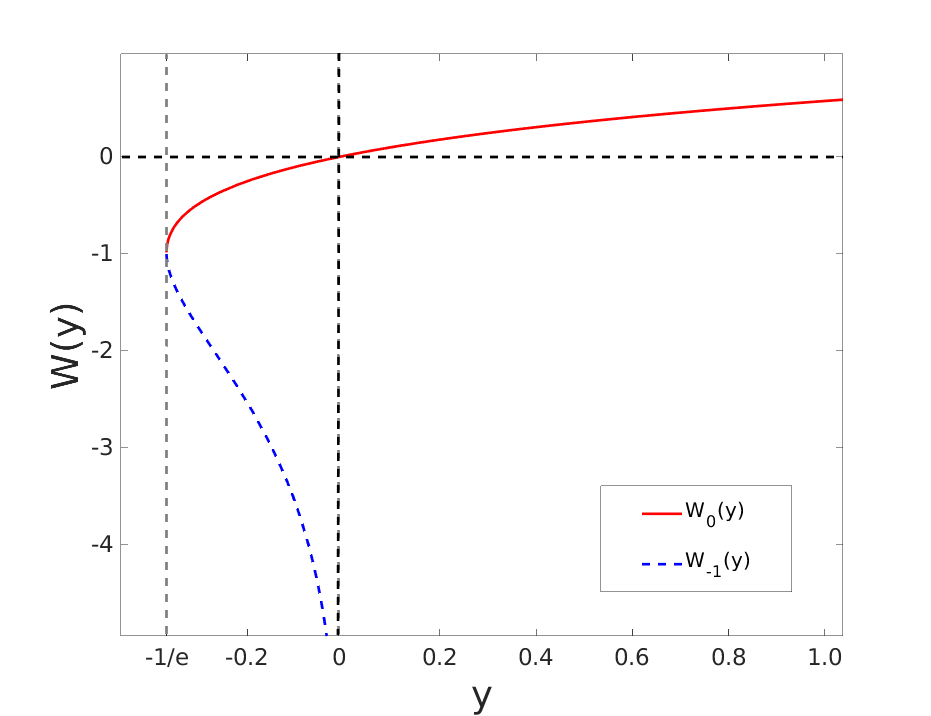}
\end{center}
\caption{Lambert function}
\label{Lambert}
\end{figure}

Lambert function has two branches in the real plane,
$W_{0}\equiv W\left( 0,y\right) $ and $W_{-1}\equiv W\left( -1,y\right) $,
which join smoothly at $W=-1$, for $y=-1/e,$ see Fig.~\ref{Lambert}. The joining point
corresponds to the sonic point ~$M=1$ located at the maximum of the magnetic
field. The upper branch $W(0,y)$,  for -$e^{-1}<y<0$ corresponds to the $M<1$ part of the
solution before the singular point  
$M\left( z\right) =\left[ -W(0,-b^{2}\left( z\right)
/e\right] ^{1/2}$.   The lower branch $W(-1,y)$, in the same range %
$-e^{-1}<y<0$,   corresponds to the $M>1$ part of the accelerating solution, $M\left( z\right)
=\left[ -W(-1,-e^{-1}b^{2}\left( z\right) \right] ^{1/2}$. 
The
function $W\left( -1,y\right) $ has a slow logarithmic divergence for $%
y\rightarrow 0-\varepsilon $ with the asymptotic $W\left( -1,y\right) $ $%
=\ln (-y)-\ln (-\ln (-y)$ \cite{}. Thus the plasma velocity outside of the
nozzle for $b(z)\rightarrow 0$ can be approximated as
\begin{equation}
M\left( z\right) \simeq \left[ -\ \ln (-y)+\ln (-\ln (-y)\right] ^{1/2}.
\end{equation}%
for $y=B^{2}\left( z\right) /\left( eB_{m}^{2}\right) \rightarrow 0$. Of
course, this solution becomes invalid when the magnetic field decreases so
that the ions can no longer be considered magnetized.

It is important to note that the solution (\ref{m}) is a truly global solution:
regularization of sonic point at $M=1$ fixes the value of the velocity
derivative at $M=1$, and the profile and the magnitude of the velocity in
the whole range from subsonic $M<1$ to supersonic region $M>1$ region. Also
note that while the density profile is also fixed,  the absolute value of the
density can be rescaled to a given value $n_{0}$ at the left boundary of the
accelerating region $0<z<L.$ 

\begin{figure}[h]
\begin{center}
\includegraphics[width=120mm]{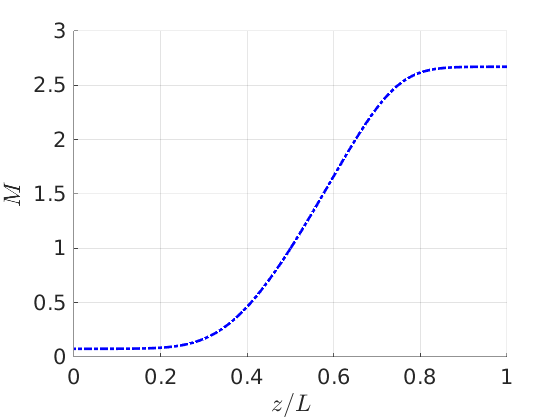}
\end{center}
\caption{Global profile of the normalized ion velocity, $M=V_\Vert/c_s$, with $M=7.56 \times 10^{-2}$ at the inlet  point, $z=0$, and $M=2.67$ at the exit, $z=L$.}
\label{M-profile}
\end{figure}

\begin{figure}[h]
\begin{center}
\includegraphics[width=120mm]{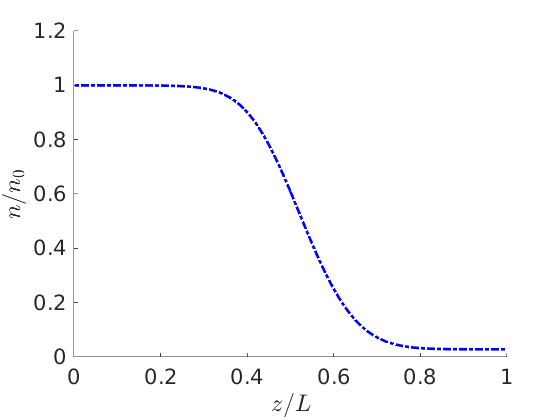}
\end{center}
\caption{Global profile of the normalized plasma density.}
\label{density-profile}
\end{figure}

\begin{figure}[h]
\begin{center}
\includegraphics[width=120mm]{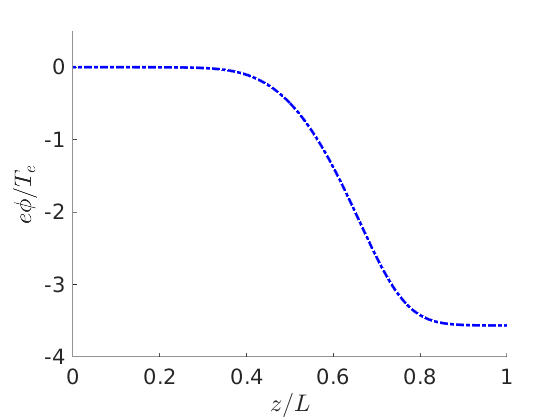}
\end{center}
\caption{Global profile of the accelerating electrostatic potential.}
\label{phi-profile}
\end{figure}

As an example, the global profiles of plasma density,
velocity, and potential are shown in Figs. \ref{M-profile}, \ref{density-profile}, and \ref{phi-profile}, 
for the magnetic field  in the
form
\begin{equation}
B\left( z\right) =\frac{B_{0}-B_{m}\exp \left( -L^{2}/4\delta ^{2}\right) }{%
1-\exp \left( -L^{2}/4\delta ^{2}\right) }+\frac{\left( B_{m}-B_{0}\right) }{%
1-\exp \left( -L^{2}/4\delta ^{2}\right) }\exp \left( -\frac{\left(
z-z_{m}\right) ^{2}}{\delta ^{2}}\right) ,  \label{b}
\end{equation}%
where $z_{m}=L/2$, $B_{0}=B\left( 0\right) ,$ the maximum magnetic field
at $z=L/2, $ $B_{m}=B\left( L/2\right)$. In what follows, the point $z=0$ will be
called the nozzle inlet, and the $z=L$ is the nozzle exit. Here, we take the mirror ratio, $R=B_m/B_0=8.04$. 

In the
example (\ref{b}), we take $B\left( 0\right) =B\left( L\right) $ for
simplicity, but this is not required. Any converging-diverging configuration
with a single maximum of the magnetic field will have a global solution
where the plasma velocity is fully determined by the magnetic field according to the equation (\ref{meq}). The
 value of the velocity at the exit of the nozzle is only defined by the
mirror ratio at the exit. Similarly, velocity at any point along the nozzle is independent of the
velocity at the inlet point but fixed by the ratio of the local magnetic field at the point to the magnetic field at the maximum. It also means that the velocity at the inlet cannot be arbitrary and is also fully defined by the ratio of the magnetic
field at the inlet  to the magnetic field in the maximum. For the magnetic field from equation (\ref{b}), with $R=8.04$,  the velocity at the entrance point $V_\Vert/c_s=(-W_{0}(-R^2/e))^{1/2}=7.56 \times 10^{-2}$, and the velocity at the exit $V_\Vert/c_s=(-W_{-1}(-R^2/e))^{1/2}=2.67$. 

In practice, experiments often show that electrons are not isothermal \cite{TakahashiPRL2018,LittlePRL2016,TakahashiPRL2020,SheehanPSST2014}. Therefore, it is of interest to generalize this analysis  for a polytropic equation of
state for electrons in the form
\begin{equation}
p_{e}=p_{0}\left( \frac{n}{n_{0}}\right) ^{\gamma }.  \label{pol}
\end{equation}%
In general, electron pressure and density can be normalized to the values at any arbitrary point,  $p_{e}=p_{0}\left( n_{0}\right)$. It is convenient
however to define the $p_0$ and $n_0$ as the values at the inlet point $%
z=0$.

Excluding the electron pressure and the electric field from equations (\ref%
{continuity2}-\ref{potential-gradient}), one obtains the following equation
for plasma velocity
\begin{equation}
\left[ V_{\Vert }^{2}-c_{0}^{2}\left( \frac{n}{n_{0}}\right) ^{\gamma -1}%
\right] \frac{\partial V_{\Vert }}{\partial z}=-c_{0}^{2}\left( \frac{n}{%
n_{0}}\right) ^{\gamma -1}V_{\Vert }\frac{\partial \ln B}{\partial z}.
\label{s0}
\end{equation}%
where $c_{0}^{2}=m_{i}^{-1}\left. \partial p_{e}/\partial n\right\vert
_{n_{0}}=\gamma p_{0}/(n_{0}m_{i})$ is the sound velocity at a point $%
n=n_{0} $. The sonic point singularity occurs at a point where the
ion velocity becomes equal to the local value of the sound velocity
\begin{equation}
c_{s}^{2}\equiv \left. \frac{\partial p_{e}}{m_{i}\partial n}\right\vert
_{s}=\gamma \frac{p_{s}}{m_{i}n_{s}}=c_{0}^{2}\left( \frac{n_{s}}{n_{0}}%
\right) ^{\gamma -1}.  \label{s1}
\end{equation}%
The solution is made regular by requesting that at the sonic point $%
\partial \ln B/\partial z=0$. The regularization near this point gives the
condition
\begin{equation}
\left( \frac{\partial V_{\Vert }}{\partial z}\right) ^{2}=-\frac{1}{\gamma +1%
}c_{s}^{2}\frac{\partial ^{2}\ln B}{\partial z^{2}}>0.  \label{s2}
\end{equation}%
Equation (\ref{s0}) with the conditions (\ref{s1}) fully define the
regular (smooth) solution across the whole acceleration region.

Equation (\ref{s0}) can be integrated, but it is more convenient to obtain
the solution directly from integrals of equations (\ref{continuity2}-\ref%
{potential-gradient}). Energy conservation gives%
\begin{equation}
V_{\Vert }^{2}=c_{s}^{2}-\frac{2e}{m_{i}}\phi ,  \label{ien}
\end{equation}%
the electron momentum balance \
\begin{equation}
e\phi =\frac{p_{s}}{n_{s}}\frac{\gamma }{\gamma -1}\left( \left( \frac{n}{%
n_{s}}\right) ^{\gamma -1}-1\right) ,  \label{iphi}
\end{equation}%
and the flux conservation
\begin{equation}
\frac{nV_{\Vert }}{B}=\frac{n_{s}c_{s}}{B_{m}}.  \label{if}
\end{equation}
Note that here the potential is measured from the sonic point, so $\phi =0$
at $z=z_{m}$ where $V_{\Vert }=c_{s}$, as  it follows from the regularization
condition at $z=z_{m}$ with $B=B_{m}$.

Excluding the potential and density, one gets an implicit equation that
defines the ion velocity in terms of the magnetic field mirror ratio at any
point inside the region $0<z<L$, with the maximum magnetic field at  $z=z_{m}$ inside the region, $0<z_m<L$,
\begin{equation}
M^{2}-1=-\frac{2}{\gamma -1}\left( \left( \frac{B}{MB_{m}}\right) ^{\gamma
-1}-1\right) .  \label{m2}
\end{equation}%
The Mach number here  is defined as the ratio of the ion local velocity to
the value at the sonic point $M=V_{\Vert }/c_{s}$. Considering that
\begin{equation}
\lim_{\gamma \rightarrow 1}\frac{1}{\gamma -1}\left( x^{\gamma -1}-1\right)
\rightarrow \ln x,
\end{equation}%
one can see that equation (\ref{m2}) for the isothermal  case $\gamma =1$ reduces to equation (\ref{meq}).

It is important to note that for general polytropic equation of state, in addition to a free density normalization parameter, which can
be taken either as the density at the entry point, $n=n_{0}$ for $z=0,$ or
the density at the sonic point $n=n_{s}$ for $z=z_{m},$ one has to introduce
the additional parameter - the electron pressure (or temperature) at the
respective reference point. Using the $z=z_{m}$ sonic point as a reference,
the value at the entry point $z=0$ can be defined as
\begin{equation}
p_{0}=p_{s}\left( \frac{n_{0}}{n_{s}}\right) ^{\gamma }.
\end{equation}%
Respectively, one can redefine the Mach number in terms of the ion velocity
to the sound  velocity $c_{0}$ at the entry point $M^{^{\prime }}\equiv
V_{\Vert }/c_{0}$.

\begin{figure}[h]
\begin{center}
\includegraphics[width=150mm]{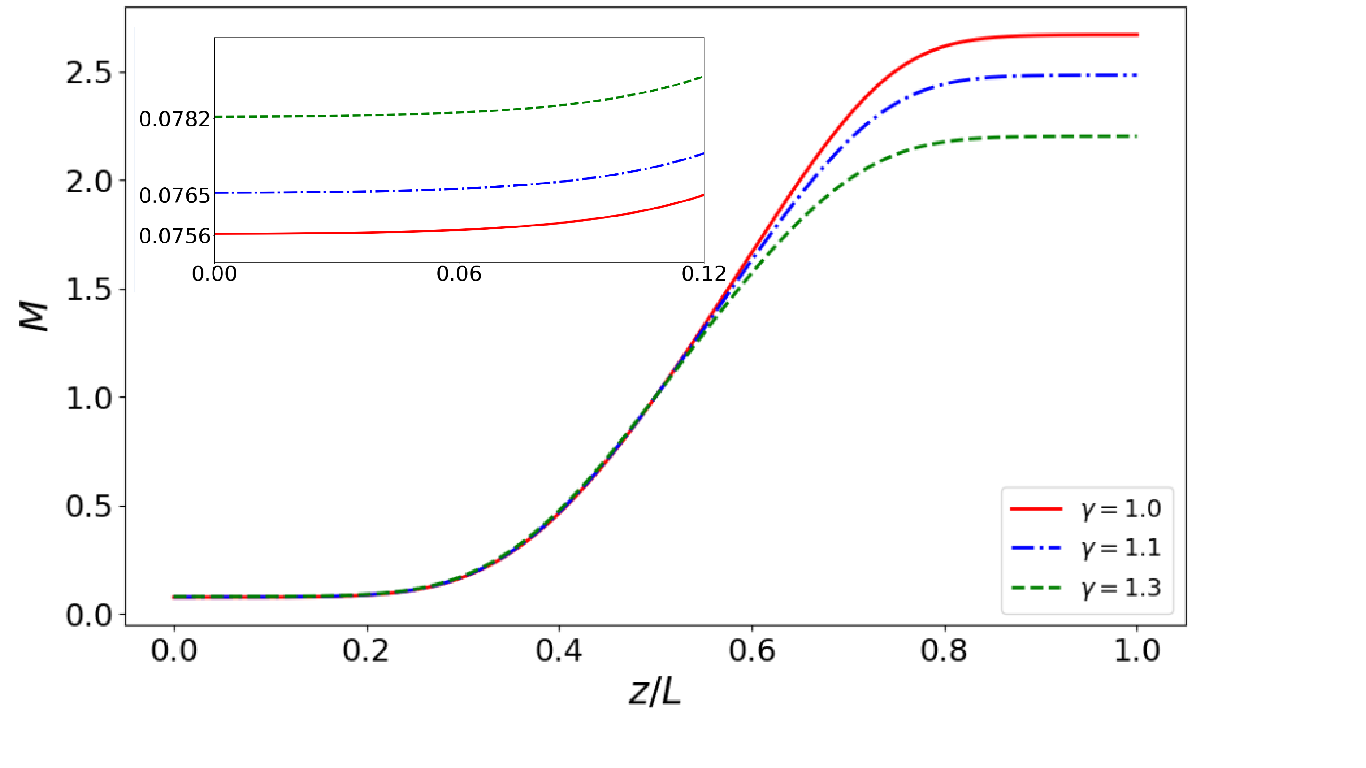}
\end{center}
\caption{Normalized ion velocity  for different $\gamma$ values.}
\label{mach_number_insert}
\end{figure}

\begin{figure}[h]
\begin{center}
\includegraphics[width=150mm]{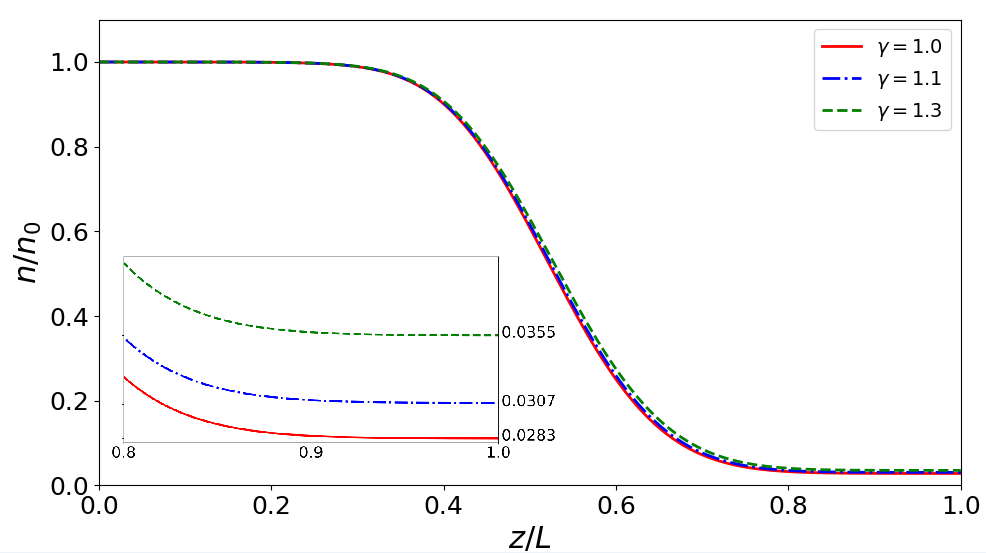}
\end{center}
\caption{Normalized plasma density for different $\gamma$ values.}
\label{plasma-density-insert}
\end{figure}

The global solutions for the ion velocity and density with the magnetic field given by (\ref{meq} are shown
in Figs. \ref{mach_number_insert} and \ref{plasma-density-insert} for different values of the polytropic coefficient $\gamma$. The solution for the velocity  is also "stiff": the full velocity profile is fully determined by the magnetic field profile. The insert in Fig. \ref{mach_number_insert} shows the values at the inlet  side of the nozzle. The density has a free normalization parameter $n_0$, but the profile is stiff otherwise.

We have shown that the magnetic barrier (converging-diverging magnetic
field) is required for the formation of the quasineutral potential structure
accelerating plasma, and presented exact solutions for the general
polytropic equation of state for electrons. An important property of such
solutions is that the normalized ion velocity at any given point is uniquely
determined by the ratio of the magnetic field magnitude at this point to the
value of the magnetic field in the maximum $B_{m}$, cf. Eqs. (\ref{m}) and (%
\ref{m2}). Such global solutions (in the whole acceleration region from the
initial value of $V_{\Vert 0}<c_{s}$ to the final exit value $V_{\Vert
}>c_{s}$) are determined by the regularization condition at the sonic point.
This also means that the finite acceleration is independent from the details
of the magnetic field profile but only determined by the mirror ratio $%
B_{m}/B_{L}$, where $B_{L}$ is the magnetic field at the apparent end of the
nozzle and therefore independent of the details of the profile inside the
barrier, for  $z<z_{m}$ as long as the magnetic field has a
maximum at $z=z_{m}$. \

In reality, the assumptions of the model become violated at low values of
the magnetic field outside of the nozzle when the ions become unmagnetized and no longer follow magnetic field lines. The detachment position depends
on the mechanisms of the detachment which are still under discussions \cite{BreizmanPoP2008,AhedoPoP2010}. The
global nature of the velocity profiles raises another question of how such
solutions are matched to the plasma source  inside the mirror region at the
entrance into the quasineutral accelerating potential structure
and how the smooth quasineutral solution can be obtained in the magnetic field with several extrema. It is worth noting, that the sonic point regularization condition also allow the smooth decelerating solution with $\partial M/\partial z <0 $.

In general, equations of our simple model should be modified to include plasma sources
to self-consistently match the source regions with the accelerating solution
which smoothly continued through the nozzle. It is worth noting that while
the value of plasma velocity is fixed at the inlet point $z=0$, the  density
profile has a free normalization parameter, e.g. the density at the inlet 
$n_{0}$. Therefore, the total plasma flux through the nozzle will be
determined by plasma density in the source, and eventually by the energy
deposited into the plasma source. Similarly, plasma thrust $T$, which is
important for propulsion applications, $T=m_{i}nV_{\Vert }^{2},$ at the exit
will be determined by plasma density in the source; effectively by plasma
pressure,  since  the electron temperature is fixed at $z=0$ and for a given equation of state. We note that in the paraxial model considered here, the additional thrust due to the plasma current induced by the 
external magnetic coil \cite{FruchtmanPoP2012} is not included, and the total thrust  is simply due to
the electron pressure.

Our results are applicable to a simple case of cold fully magnetized ions. 
The effects of finite ion pressure also contribute to the ion exhaust
velocity (and therefore to the thermal pressure generated thrust) via the
mirror force but are neglected in our model here. Additional forces (such as due to the ionization or collisions) in the momentum balance, as well as geometrical expansion effects, will shift the position of the sonic point. With additional forces it is also possible to have a smooth sonic point transition without the maximum of the magnetic field \cite{ManheimerIEEE2001,CohenZurPoP2002,SmolyakovAIAA2019}, however  the resulting accelerating potential profiles remain global, i.e. stiff, with a similar property of the unique solution defined by the sonic point regularity condition.  This property (based on the formally similar equations for the acceleration in the Laval nozzle)  is generally shared by a wide class of gasdynamics systems, Hall thrusters \cite{CohenZurPoP2002,AhedoPoP2001}, and magneto plasma dynamics systems where plasma is accelerated by the magnetic pressure\cite{FruchtmanPoP2003}. 

In general,  kinetic effects of the electron and ion trapping should be included, as well as  dissipative processes such as heat fluxes, charge-exchange interactions with neutrals, and
ionization. Such effects might be important  in fusion applications of the magnetic expanders \cite{RyutovAIP2016,TogoNME2019,OnofriPoP2017} and in propulsion applications (for  a recent overview of the physics of the magnetic nozzle for propulsion applications see Ref. \onlinecite{KaganovichPoP2020}).
Presence of high energy species, and coupling of  plasma expansion with  the Debye length phenomena and non-quasineutral effects of classical  double layers structures (in the sense of Ref. \onlinecite{PerkinsPRL1981}) bring further complications and several different scenarios for the formation of the accelerating potential structures \cite{AhedoPRL2009}.  An interesting question is a possibility of the formation of the Debye layer at the sonic point analogous  to the weak shock solutions  in gas dynamics.  

Nevertheless, despite a number of simplifications, the solutions presented here provide useful insight  on the mechanism for formation of accelerating potential structures in the magnetic mirror configurations. These results
provide a simple illustration to seemingly surprising experimental results in VASIMR \cite{LongmierPSST2011} that
did not find narrow Debye type double layer but show the wide accelerating potential structures. According to the physical picture presented here, such
accelerating structure occurs due to presence of the magnetic barrier in
the converging-diverging magnetic field. These results should also be
useful for the tests and benchmarking of numerical simulations \cite{BaalrudPoP2011,ChenPRE2020} and interpretations of the results from more complete models.

\begin{acknowledgements}

This work was supported in part by NSERC Canada and the U.S. Air Force Office of Scientific Research FA9550-15-1-0226 and FA9550-21-1-0031. Computational resources were provided by Compute Canada.
A.S. would like to acknowledge illuminating discussions with S. I. Krasheninnikov and Y. Raitses. 

\end{acknowledgements}

 \section*{Data availability} No new data was generated in this study.  



\bibliographystyle{unsrt}
\bibliography{References}

\newpage

\end{document}